\documentstyle[psfig]{article} 
\title{Vibrational decoherence in ion-trap quantum\\
computers
\thanks{E-mail:agarg@nwu.edu;
To appear in Proceedings of SPIE, The International Socity for Optical Engineering,
Vol. 3385; AeroSense '98, Orlando, FL April 13-17, 1998.}} 

\author{Anupam Garg\\
Department of Physics and Astronomy,\\
Northwestern University,\\
Evanston, IL 60208, USA}

\begin{document} 
\maketitle 

\def\al{\alpha}
\def\be{\beta}
\def\om{\omega}
\def\dta{\delta}
\def\Dta{\Delta}
\def\sig{\sigma}
\def\ul{\underline}
\def\ket#1{|#1\rangle}
\def\bra#1{\langle #1|}
\def\gtoe{{\it g} \leftrightarrow {\it e}}
\def\tvib{\tau_{\rm vib}}
\def\trad{\tau_{\rm rad}}
\def\beq{\begin{equation}}
\def\eeq{\end{equation}}
\def\bea{\begin{eqnarray}}
\def\eea{\end{eqnarray}}
\def\nnu{\nonumber}
\def\vla{\relbar\joinrel\relbar\joinrel\relbar\joinrel\longrightarrow}
\def\Bp{B_\perp}
\def\cvim{{\vec{{\bf c}}}_{i\mu}}
\def\cvin{{\vec{{\bf c}}}_{i\nu}}
\def\amp{{\cal A}}
\def\ham{{\cal H}}
\def\hqc{\ham_{\rm QC}}
\def\hnm{\ham_{\rm nm}}
\def\Snm{S_{\rm nm}}
\def\dint{\int\!\!\!\int}
\def\tint{\dint\!\!\!\int}
\def\fint{\tint\!\!\!\int}
\def\bvf{{\bf f}}
\def\bvB{{\bv B}}
\def\Fcnn{F^{\rm Coul}_{\rm nn}}  

\begin{center}
{\bf Abstract}
\end{center}
Decoherence is studied in an attractive proposal for an actual implementation
of a quantum computer based on trapped ions. Emphasis is placed on the decoherence
arising from the vibrational motion of the ions, which is compared with that due
to spontaneous emission from excited states of the ions. The calculation is made
tractable by exploiting the vast difference in time scales between the vibrational
excitations and the intra-ionic electronic excitations. Since the latter are several
orders of magnitude faster, an adiabatic approximation is used to integrate them
out and find the inclusive probability $P(t)$ for the electronic state of the ions
to evolve as it would in the absence of vibrational coupling, and the ions to evolve
into any state whatsoever. The decoherence time is found at zero temperature and
for any number of ions $N$ in the computer.  
Comparison is made with the spontaneous emission decoherence, and the implications
for how trap voltages and other parameters should be scaled with $N$ are discussed.

%\keywords{quantum coherence, ion traps, adibatic approximation,
%spontaneous emission}

\section{INTRODUCTION}
\label{sect:intro}

A quantum computer (QC) is a proposed machine that would make use of the superposition
principle of quantum mechanics to do certain types of calculations with an
unprecendented degree of parallelism. The concept of a quantum computer goes back
to the early 1980's.\cite{Ben,Fey}
It is an outgrowth of the view that any computation is
ultimately a physical process. This view has tended to get obscured in recent times,
especially as our understanding of the universal mathematical properties of
computation and associated issues in complexity and computational theory has grown.
If one thinks about the computer as a machine, however, in which wheels must turn, or
electrons must be transported between transisitors, and so on, then it is somewhat
surprising, at least in hindsight, that it has taken us so long to start thinking about
the implications of the laws of quantum mechanics for the turning of the wheels,
transport of charges etc. To avoid misunderstanding it should be emphasized that
quantum mechanics already plays a vital role in the functioning of all modern electronic
circuitry, in that it underlies the concepts of energy bands,
the effects of doping on semiconductors, the theory of conductivity, the nature of
electron and hole statistics, and many other aspects of semiconductor and device physics.
The ultimate equations that we use to describe circuits, such as Ohm's law, or Shockley's
equation, are deterministic and classical. Quantum mechanics provides the microscopic basis
for these equations, and fixes the parameters appearing in them. To {\it use} these
equations, however, one only needs to know the meaning of the terms which appear in them,
and the scope of their applicability.

In this introductory section we will describe what a quantum computer is. In Sec.~2,
we shall describe the ion-trap quantum computer proposed by
Cirac and Zoller\cite{cz}, and summarise the results of an earlier
calculation\cite{agprl} of the decoherence time in this device. An approximate
account of this calculation, less technical than that in Ref.~\cite{agprl} is
given in an Appendix, along with some details about the structure of the ion array in the
trap. Since many reviews of quantum computers now exist,\cite{Jason} this introductory
section may well be redundant for many readers. Readers who are unfamiliar with the
subject or with the terms used to describe the contents of the rest of the paper
may, however, profit from it.

\subsection{What is a Quantum Computer?}

The key point is that the state of a classical computer is definite at all times. This
is the prevailing rule in classical mechanics. In quantum mechanics by contrast, the state
of a physical system is not definite, and a system does not have definite properties
until they are measured. The well known uncertainty principle, which expresses the
impossibility of simultaneously knowing both the momentum and the position of a particle
with definiteness, i.e., with zero uncertainty, is an expression of this fact. In general,
therefore, the state of a quantum system must be expressed as a linear superposition of
some basis states. Time evolution preserves this superposition property, and since
it also conserves probability, must be described by a unitary operator, as long as the
system is not externally disturbed.

To understand how these ideas can be usefully applied to make a computer, it is best to
take a simple example, that of evaluation of an integer valued
function of an integer argument. The
first point has to do with the correspondence between numbers and physical states. Thus
a single bit (0 or 1) is represented by a physical system or register
which can exist in two states,
which are represented by $\ket 0$ and $\ket 1$. (There are many examples of a two-state
system of which we mention only two: the spin of a particle of spin-1/2, and the electronic
states of an atom or ion. In the latter case, there are un general many more than two states,
and the two-state description is only an approximation. One must be careful to avoid 
perturbations which will excite the atom or ion into other states.) The novel aspect of
a quantum system is that where a classical bit must either be a 0 or a 1, a quantum bit
can be any suerposition
\beq
\ket\psi = C_0 \ket0 + C_1 \ket1, \label{psi01}
\eeq
where $C_0$ and $C_1$ are arbitrary complex numbers, subject to a normalization condition
such as $|C_0|^2 + |C_1|^2 = 1$, expressing the fact that the system has unit total
probability for being in some state or other. (We shall leave this normalization out
below.  This keeps the formulas looking neat, and it is easy to reimpose at any point
in the argument.) 

The state (\ref{psi01}) has no classical interpretation. In particular it is simply
incorrect to say that the system {\it is} in one of the states $\ket0$ or $\ket1$ with
certain probabilities, in the way that we can say that a coin which we toss with our
eyes closed {\it is} either heads or tails even though we don't know exactly which one
it is. Such an
interpretation of the state leads to wrong predictions for the outcome of many
measurements. Similarly, saying that the system is simultaneously in states $\ket0$
and $\ket1$ has only limited usefulness. One simply has to accept (\ref{psi01}) as
the state of the system, and use the quantum mechanical assignment of measurables to
operators to make predictions.

Next, let us consider a quantum register composed of two two-state systems.
The states of this register
can be written as $\ket{0\,0}$, $\ket{0\,1}$, $\ket{1\,0}$ and $\ket{1\,1}$. Here
$\ket{1\,0}$ means that the first two state system is in state 1 and the second in state 0.
These four states can be regarded as representations of the four binary numbers 00, 01,
10, and 11. Again, as in Eq.~(\ref{psi01}), our two two-state-system register can be in a
superposition
\beq
\ket\psi = C_{00} \ket{0\,0} + C_{01} \ket{0\,1}
           + C_{10} \ket{1\,0} + C_{11} \ket{1\,1}.
\label{psi4}
\eeq
Generalizing to a register of $N$ two-state systems, it is obvious that one can represent 
$2^N$ numbers in this way. The general superposition can now be written as
\beq
\ket\psi = \sum_{n=0}^{2^N -1} C_n \ket n,
\label{psin}
\eeq
where we use the notation $\ket n$ to represent a state in which the binary representation
of the number $n$ yields the individual state of each two-state system.

Suppose now that we wish to evaluate an integer valued
function $f(n)$ of the integers. (We suppose that both $n$ and $f(n)$ are bounded so that
we can represent them with registers of managable size.) 
Let us assume that we have found a means of dynamically evolving
our system in such a way that the corresponding unitary operator $U_f$ transforms the
state $\ket n$ into $\ket{f(n)}$. More precisely, it is necessary to consider a computer
with two registers, which we call the input and the output. The initial state of our
computer is taken to be
\beq
{\ket\psi}_{\rm initial} = {\ket n}_{\rm in} {\ket 0}_{\rm out}.
\label{psiin}
\eeq
Then the unitary operator $U_f$ is supposed to be such that
\beq
{\ket\psi}_{\rm final} = U_f {\ket\psi}_{\rm initial}
                        = {\ket n}_{\rm in} {\ket{f(n)}}_{\rm out}.
\label{psif}
\eeq
Note that the input register is left unchanged, while the output register contains the
value of the function. Whether or not the operator $U_f$ can be constructed depends on
the function $f$ under consideration, and is a separate question, belonging to the theory
of algorithms and computability.

The core concept in quantum computing is to take advantage of the superposition
principle. It follows from the linearity of the unitary time evolution operator $U_f$
that
\beq
U_f \sum_n C_n {\ket n}_{\rm in} {\ket 0}_{\rm out} = 
    \sum_n C_n {\ket n}_{\rm in} {\ket{f(n)}}_{\rm out}. 
\label{qpar}
\eeq
Equation (\ref{qpar}) contains the essence of quantum parallelism. Since the effort
required to create or implement the evolution $U_f$ is separate and independent of the
state of the quantum computer, it follows that if we can prepare the initial state of the
quantum computer in the superposition on the left hand side of Eq.~(\ref{qpar}),
the final state will contain, albeit as a superposition, the function $f(n)$ for {\it all}
values of $n$. We have performed, at first sight, $2^N$ function evaluations at one shot.

\subsection{The Read-Out Problem and Shor's Breakthrough}

The parallelism of Eq.~(\ref{qpar}) is somewhat hollow, however. To obtain information on
the function values $f(n)$, one must measure the state of the input and output registers.
Such a measurement can only yield {\it one} value of $f(n)$. We have no way of knowing
the other function values. Worse, the value of $n$ for which the function is evaluated is
itself random: the probability of getting $n$ is $|C_n|^2$. Thus one could not even generate
a table of $f(n)$ by repeated runs of our machine without substantial redundancy.

The above difficulty, also known as the read-out problem, stymied the subject of quantum
computers for the decade after its genesis. Interest in the field was primarily in terms of
the theory of computability and complexity,\cite{d85,bv} and the thermodynamics of
computation, specifically the issues of reversibility and energy consumption.\cite{lan,chb}
The breakthrough came in 1994 with the discovery by Coppersmith and Shor of a quantum
algorithm for factorization of a composite number with two prime factors.\cite{dc,shor}
The key realization is that for certain problems what one needs is not the individual
function values $f(n)$, but some global property of the function. In the factorization
problem, the property which is exploited is the period of a periodic function. This can
be found using Coppersmith's quantum Fourier transform.\cite{dc} We refer readers
to a recent review of the number theory underlying these algorithms,\cite{ej} but briefly
and heuristically speaking, quantum mechanics is intrinsically well set up for the efficient
addition of terms multiplied by phase factors required in finding a Fourier transform.  
(Witness the closely related examples of Fraunhofer and $N$-slit diffraction in optics.)
In this way, one can indeed exploit quantum parallelism to do certain calculations very
effectively on a quantum computer. For a number with $L$ decimal digits, Shor's
algorithm factorizes it in order $L^3$ steps.\footnote{For completeness, it should be noted
that Shor's algorithm is probabilistic, in that it requires choosing a random number
for its operation. The number theoretic method employed by the algorithm fails to give
a useful result for certain choices of this random number. However, it can be shown that
the probability of success can be made to approach arbitrarily close to unity with
$O(L)$ choices of the random number. Thus this consideration does not alter the polynomial
time nature of Shor's method.} The best currently known classical algorithms, by contrast, 
require $\sim\exp[cL^{1/3}(\ln L)^{2/3}]$ steps, where $c$ is a constant of order unity.
Needless to say, this discovery has galvanized the subject. The difficulty of factorization
is the basis for much modern day cryptography, and the elimination of this difficulty would
have obvious repurcussions. In addition, it has been surmised that a generic quantum computer
may be used for the simulation of a general many body quantum mechanical problem.\cite{Jason}

\subsection{Primitive Quantum Gates}

It remains to say something about how the unitary operator $U_f$ is to be constructed.
One would like to do so by composing a small number of basic or primitive
operations operating on a small number of bits at a time. Indeed, 
a one-bit gate consisting of a rotation, and a two-bit gate called the controlled-not (or
C-NOT) suffice to build any operator $U_f$, in the same way that an AND and a NOT suffice
for a classical computer. (Equivalently, one could say that any unitary
matrix acting on an $n$ dimensional space can be built by multiplying $n(n-1)/2$ unitary
matrices that act on two-dimensional subspaces at a time. The general point that such simple
operations can be composed to form arbitrarily complex operators is obvious to any one
who has ever solved linear equations by Gaussian elimination or solved matrix
problems numerically, but readers who wish to see this codified as theorems should consult
Refs.~\cite{d85,ddv,bdej,sw}. The physical point which is important is that it is necessary
to have a means of introducing correlations amongst the states of two different two-state
systems, and this is done via the controlled-not. Any other two-bit gate that was not
too close to the identity operation would work just as well.)
The one-bit rotate acts on a single bit as follows:
\beq
U(\phi)(C_0\ket 0 + C_1\ket 1)  =  (C_0\cos\phi + C_1\sin\phi) \ket0 
                         + (-C_0\sin\phi + C_1 \cos\phi) \ket 1,
\label{rot}
\eeq
while the C-NOT acts on any pair of bits as per the truth table in Table 1.
\begin{table} [t]   %>>>> [h] means place table here
\caption{Truth Table for the Controlled-Not Gate
\label{tab1}}
\vspace{0.2cm}
\begin{center}       
\begin{tabular}{|c|c|} 
\hline
\rule[-1ex]{0pt}{3.5ex}  Input & Output  \\ %% \rule[]{}{} opens up each row
\hline
\rule[-1ex]{0pt}{3.5ex}  $\ket{0\,0}$ & $\ket{0\,0}$ \\
\rule[-1ex]{0pt}{3.5ex}  $\ket{0\,1}$ & $\ket{0\,1}$ \\
\rule[-1ex]{0pt}{3.5ex}  $\ket{1\,0}$ & $\ket{1\,1}$ \\
\rule[-1ex]{0pt}{3.5ex}  $\ket{1\,1}$ & $\ket{1\,0}$ \\
\hline
\end{tabular}
\end{center}
\end{table}
The action of this gate is best described in words. It leaves the first (or control) bit
unchanged, while the second (or target) bit is flipped from 1 to 0 or vice versa
if the control bit is a 1, and left unchanged if the control bit is a 0.
It should of course be remembered that this operation also acts on linear
superpositions. Thus,
\beq
a \ket{0\,1} + b \ket{1\,1} \quad \buildrel \rm C-NOT \over \vla \quad
a \ket{0\,1} + b \ket{1\,0}.
\label{cnotex}
\eeq
It should be noted in this connection that the C-NOT is defined
with respect to a very definite choice of basis states, which is referred to as the
computational basis. It should also be noted that in practical implementations,
it may be more useful to construct a few more gates than just the C-NOT and the one-bit
rotate. Examples of such gates and simple circuits like adders and multipliers may be
found in Refs.~\cite{Fey,Jason}.

The above discussion of the construction of the evolution operator $U_f$ is very
general, and does not say how the individual gate operations are to be carried out.
There are at least two broadly different types of quantum computers that have been
discussed in the literature. In Feynman's original conception\cite{Fey}, the idea
is akin to having an array of sites between which an electron can move, and to arrange
the terms in the Hamiltonian to be such that matrix elements between different sites
implement the desired gates. It is important that the connections be local in order
that the gates remain simple.  A computation then proceeds in analogy with an electron
wave packet coursing through the sites. The second conception, which is the one more
commonly discussed today is to have a temporal sequence of gate operations applied
through external perturbations to a fixed set of quantum bits.

With this somewhat long preamble, we now turn to the physical implementation of a
quantum computer, especially that based on ion traps.

\section{THE ION-TRAP QUANTUM COMPUTER}
\label{sect:czqc}

The desiderata in making a real quantum computer are almost too obvious to state. One
needs first and foremost to have a means of controlling and driving the bits individually.
Secondly, one needs to have low dissipation, or interaction with the environment. Equation
(\ref{qpar}) assumes that the computer forms a closed system, isolated from the rest of
the world. In reality, interactions with the environment will end up entangling the
state of the computer with that of the environment. These interactions are most
likely to be such that the correlations amongst the computational degrees of freedom (or
quantum bits) get lost over time. This point is well known from the study of problems
like NMR, where also one has many quantum systems, each with a Hilbert space of small
dimensionality, but interacting with an environment --- the lattice, motional degrees of
freedom of the atoms, electromagnetic radiation, etc.

Even after the above conditions are met, it is not easy to implement a good two-bit gate. 
In addition to the ion-trap proposal\cite{cz} that we shall describe in more detail,
another one based on caity QED\cite{pgcz} seems promising. In fact, experimental two-bit
gates, inspired by each of these schemes, were demonstrated shortly after their
proposal.\cite{cm,qat}

The ion-trap QC utilizes an array of $N$ identical ions in a linear rf Paul trap, each
of which can be independently addressed and driven by a laser. (See Fig. 1.)
\begin{figure}
\centerline{\psfig{figure=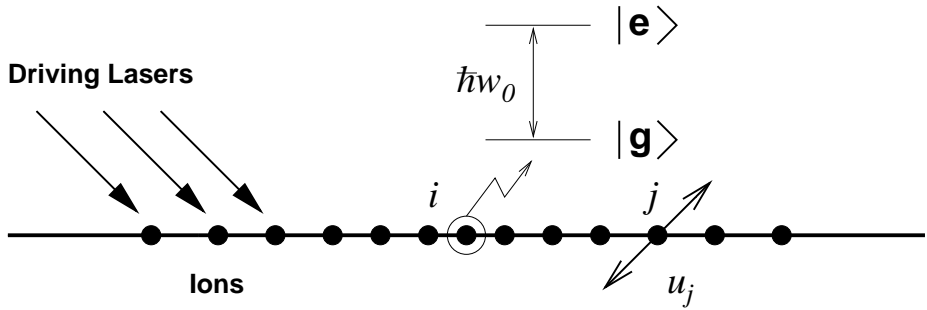}}
\caption{Schematic of the ion-trap quantum computer.}
\end{figure}
The ions are cooled to nearly zero temperature.
Such arrays have been studied for some time now
because of the opportunities they offer for spectroscopy, and developing frequency
standards.\cite{mgr,djb} The ions have a nonuniform spacing determined by the Coulomb
repulsion between the ions and the effective trap potential. A quantum bit is composed
of two internal states of an ion, which we denote by $\ket e$ and $\ket g$, and one-bit
gates are executed by applying suitable pulses ($\pi/2$, $\pi$, etc.) to the ions.
The clever idea which enables the execution of a two-bit gate is to use the center
of mass motion of the entire array as an additional computational degree of freedom.
Let $\om_0$ be the $\gtoe$ transition frequency of the ions, and let $\om_z$ be the
center of mass vibrational frequency. A laser pulse with carrier frequency 
$\om_0 \pm \om_z$ acting on the {\i\/}th ion entangles the internal state of this ion
with that of the center of mass mode. As discussed in Ref.~\cite{cz}, a sequence
of three such pulses, the first and third applied to ion {\it i}, and the second to ion
{\it j\/}, have the net effect of a two-bit gate in which ions {\it i\/} and {\it j\/} are
the control and target bit, respectively. In fact, by also applying one-bit rotate
operations to ion {\it j\/}, this gate can be exactly turned into the C-NOT.
A sequence of one- and two-bit gate operations can then be executed, yielding any
desired computation, as discussed in Sec.~\ref{sect:intro}.

\subsection{Spontaneous Emission Decoherence}

The advantages of the ion-trap QC are its conceptual simplicity, and high degree of
isolation from the environment. In fact, the proposal is regarded seriously enough
that an effort is underway to build a proof-of-principle prototype.\cite{rjh}
The most obvious source of decoherence is the spontaneous emission from the ions. It
seems reasonable that effects like superradiance, in which the proximity of the ions
to one another affects this decay, can be neglected, and that each ion can be treated
separately. If we assume that half the ions are in an excited state at any given time
on the average, then in a computer with $N$ ions, any computation must be completed
in a time of order $\trad = 2\tau_s/N$, since even one spontaneous emission event
destroys the phase coherence among the computational basis states of the QC as a whole.
One can minimize this source of decoherence by employing ions where the $\ket e$ state
can not decay to any state other than $\ket g$ and the decay in question is dipole
forbidden. For example, in Ba$^+$ ions, the decay between the $5d\ ^2D_{5/2}$ multiplet
(one member of which serves as $\ket e$) and the $6s\ ^2S_{1/12}$ multiplet (which
supplies the $\ket g$ state) is via an E2 process, for which
$\tau_s = 30$-70 s.\footnote{Other ions in groups IIA and IIB, such as
Ca$^+$ and Hg$^+$, have similar level schemes, and long spontaneous decay times.} 
The drawback, as noted by Plenio and Knight\cite{pk} is, that by the general relation
between Einstein A and B coefficients, the coupling of the laser to the $\gtoe$
transition is also weak, and the time required for all gates is correspondingly long.
One can not shorten this time without limit by increasing the laser power without
starting to excite the ion into higher levels, or even ionize it further, via two-photon
absorption. The radiative decoherence can be decreased by working with a $\Lambda$
system and using Raman pulses, but it can not be totally eliminated.\cite{pk}

\subsection{Vibrational Decoherence}

The second source of decoherence in the ion-trap QC is the ionic vibration. To understand
this, suppose (see Fig. 1), ion {\it j\/} is displaced from its equilibrium position. This
creates an excess electric field or electric field gradient at any other ion, say
{\it i\/}. The time evolution in the $\ket e$, $\ket g$ space of ion {\it i\/} is thereby
altered, and in the long run, the QC is not in the intended state, i.e., we have
decoherence. Let us denote the decoherence time due to this process by $\tvib$. A proper
calculation of this time is not so easy as that of $\trad$ as all the ions are coupled
to the same bath, and can not be treated individually. A naive
consideration based on only a few ions leads to an overestimate, and is
misleading. The calculation presented in the Appendix shows that $\tvib^{-1}$ scales
more rapidly than $N$, where $N$ is the number of ions. More precisely, we find,
\beq
\tvib^{-1} \sim N^{1/2} {q^2 Q^2 \over 2\pi\hbar m\om_0\om_t s_0^8},
\label{tvib1}
\eeq
where $Q$ is a quadrupole transition matrix element (appropriate to the Ba$^+$ example),
$\om_t$ is a typical transverse normal mode frequency for the ion array, $q$ and $m$ are
the ionic charge and mass, and
$s_0$ is the minimum spacing between the ions which occurs at the center of
the array. We can write $\tvib^{-1}$ more explicitly as a rate by noting that
$q^2 = m\om_z^2d_0^3$, where $d_0$ is the the trap length scale parameter [see
Eq.~(\ref{d0})], and that
\beq
Q^2 \propto  \hbar /\tau_s k_0^5,
\label{quadmom}
\eeq
where $k_0 = \om_0/c$. It follows that
\beq
{1\over \tvib} \sim {N^{1/2} \over \tau_s} \biggl({d_0 \over s_0} \biggr)^3
                 {\om_z^2 \over \om_0\om_t} {1 \over (k_0 s_0)^5}.
\label{tvib2}
\eeq

\subsection{Discussion of Results}

Equation (\ref{tvib2}) has interesting implications for the scaling of $\tvib$ with
$N$. We can not naively take this as $N^{1/2}$ because the behaviour of $\om_z$,
$\om_t$, and $s_0$ as $N$ is increased, depends on how the trap operating conditions
are varied. Suppose $s_0$ is held fixed as $N$ is increased. Then
$(d_0/s_0)^3 \sim N^2/\ln N $ and $\tvib^{-1} \sim N^{5/2}/\ln N$. In this case,
however, the longitudinal voltage on the trap electrodes, which is proportional
to $\om_z^2$, varies as $(\ln N)/N^2$. Since the time needed for a two-bit gate
varies as $\om_z^{-1}$, the total computational time goes up. Secondly, the
longitudinal confinement becomes weaker, and non-linearities in the trapping potential,
and electrode patch voltages become more important.  Suppose, on the other hand, that
the trap voltages, and therefore, $\om_z$ and $\om_t$, are held fixed as $N$ increases.
Then $\tvib^{-1} \sim N^{35/6}(\ln N)^{-8/3}$. [The variation is as
$N^{9/2}(\ln N)^{-2}$ for an E1 transition.] Now, however, the minimum inter-ion
spacing $s_0$ varies as $\sim N^{-2/3}$. This may make it difficult to optically
resolve and address individual ions, which is basic to the operation of the QC.
It seems likely that some compromise between these two extremes will have to be sought,
depending on engineering considerations.  A general point is worth noting in this
connection.  Since the radiative and vibrational decoherence processes are independent,
the total decoherence time of the computer is given by adding their {\it rates}:
\beq
t_d = \left(\trad^{-1} + \tvib^{-1} \right)^{-1}.
\label{tdec}
\eeq
$t_d$ is the useful window of time in which any computation must be finished. Thus
if one of the two rates turns out to be much larger than the other, we can relax the
design considerations on the smaller rate, and focus on ways to reduce the larger one.

Let us estimate $\tvib$ using the example of Ba$^+$ ions and the levels mentioned
above. The frequency $\om_0 = (2\pi)1.7\times 10^{14}$ Hz. We take
$\om_z/2\pi = 100\,$kHz, and $\om_t/2\pi = 20\,$MHz. This yields  $d_0 = 14\,\mu$m.
For $N=1000$, we obtain $\tvib \simeq 10^4 \tau_s$, which is surprisingly large.
(It is even larger in comparison to $\trad = \tau_s/N$.) The drawback is that
$s_0 \simeq 0.5\,\mu$m with the same parameters. This runs into the difficulty with
optical resolution mentioned above. We have not explored compromise variations
of trap parameters in detail.

It is natural to ask if we should not have anticipated $\trad$ being so much larger
than $\tvib$. The non-trivial scaling with $N$ and the fact that the relationship would
reverse for larger $N$ makes us believe that the answer is no.

\subsection{Conclusion}

In summary, vibrational decoherence is not a significant problem in the ion-trap
QC for $N \le 10^3$, as originally envisaged by Cirac and Zoller.\cite{cz}
This optimistic result should be tempered somewhat however. The technical difficulties
in working with $10^3$ ions are enormous, and the spontaneous emission decoherence
itself is nothing to sneeze at. It is possible that the $\Lambda$ system with Raman
transitions will mitigate this difficulty, but this needs to be explored more fully.
A qualitative argument suggests that the adiabatic suppression of $\tvib^{-1}$
will not be as effective in this case, however. If this turns out to be true, it would
provide a nice example of the compromise between different types of decoherence discussed
above. 

\appendix
\section{CALCULATION OF VIBRATIONAL DECOHERENCE TIME} 
\label{sect:vdec}

We give here an approximate calculation of the vibrational 
decay time. The calculation is based on Ref.~\cite{agprl} which should be
consulted for a more precise description. 

We denote the equilibrium axial position of the {\it j\/}th ion by $z_j$, and 
its deviation from equilbrium by ${\bf u}_j$. We will consider a
$\gtoe$ transition of electric quadrupole (E2) type, and denote the ionic charge by
$q$, and the quadrupole transition matrix element by Q, ignoring all vector and
tensor indices, here and in what follows. The coupling between the vibrations
and the $eg$ space of the {\it i\/}th ion is described the Hamiltonian
\beq
V_i = \sum_{j\ne i} {qQ\over |z_i - z_j|^4} u_j
               \left(\ \ket e\bra j +{\rm h.c.} \right).
\label{pert}
\eeq

The effect of this perturbation is small for two reasons. First, its magnitude is
small, $|V_i| \ll \hbar\om_0$. Second, the time scale for its variation is
determined by the ion array's normal mode frequencies, which are of order 100 MHz at most,
and thus much less than $\om_0$, which is an optical transition frequency of order
100 THz or more. Formally, 
\beq
\left| {\dot V}_i\over V_i \right| \ll \om_0.
\label{slow}
\eeq
The perturbation is thus slow, and the internal state of the ions can follow the
vibrational disturbances adiabatically. If this following were perfect, we would have no
real excitations or transfer of energy. Decoherence arises solely because of corrections
to the adiabatic approximation, and we can thus expect the final answer for $\tvib^{-1}$
to contain a small factor like $(\om_v/\om_0)^a$ where $a$ is a positive exponent, and
$\om_v$ is a typical vibrational time. We now turn to estimating this effect.

\subsection{Adiabatic Approximation}

Let us map each two-state ion onto an equivalent spin-1/2 system,
with $\ket e$ and $\ket g$ being the up and down spin states. The {\it i\/}th spin sees
effective magnetic fields $B_z = \hbar\om_0$
and $\Bp(t) = V_i(t)$ (see Fig. 2), where we 
take $V_i(t)$ to be a specified time dependent c-number perturbation.
\begin{figure}[b]
\centerline{\psfig{figure=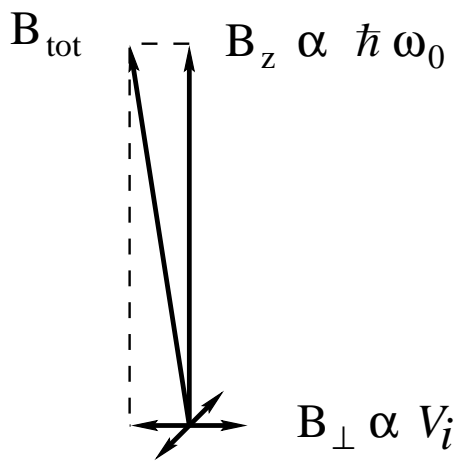}}
\caption{Equivalent magnetic field description of {\it i\/}th ion.}
\end{figure}
The total Hamiltonian for this ion can thus be written as
\beq
\ham_i = {1\over 2}\hbar\om_0 \sig_z + \hbar{\bvf}(t)\cdot{\vec\sig},
\label{hami}
\eeq
where ${\bvf}(t) = (B_x,B_y,0)/2\hbar$, and the $\sig$'s are the usual Pauli matrices
in the equivalent spin space. Note that $\bvf$ has no $z$ component.

We wish to study the time evolution of the spin for a general $\bvf (t)$ which is small
and slow. To this end, let us write a general spin state as
\beq
\ket{\psi(t)} = u_+(t) e^{-i\om_0t/2}\ket+  + u_-(t) e^{i\om_0t/2} \ket{-}.
\label{psispin}
\eeq
where $\ket\pm$ denotes the eigenstates of $S_z$ with eigenvalues $\pm 1/2$.
Schr\"odinger's equation gives
\beq
i{\dot u}_{\pm} = e^{\pm i\om_0 t} f_{\mp}(t) u_{\mp}(t),
\label{udot}
\eeq
where $f_\pm = f_x \pm i f_y$.  Since $f_\pm$ varies very slowly, we try a solution
to Eq.~(\ref{udot}) of the form
\beq
u_\pm(t) = \al_\pm(t) +\be_\pm(t),
\label{upm}
\eeq
where $\al_\pm$ and $\be_\pm$ are functions that vary rapidly and slowly, respectively.
The latter is almost a constant over a period $2\pi/\om_0$, and the former almost averages
to zero.

We now substitute Eq.~(\ref{upm}) in Eq.~(\ref{udot}), and separately equate the fast
and slowly varying parts. This yields
\begin{eqnarray}
i{\dot \al}_{\pm} & = & e^{\pm i\om_0 t} f_{\mp}(t) \be_{\mp}(t), \label{abdot1} \\
i{\dot \be}_{\pm} & = & e^{\pm i\om_0 t} f_{\mp}(t) \al_{\mp}(t).
\label{abdot2}
\end{eqnarray}
The next step is to integrate these equations. Consider Eq.~(\ref{abdot1}) first.
To a first approximation, we can regard $f_\pm$ and $\be_\pm$ as constants. Integration
then gives
\beq
\al_{\pm} = \mp \om_0^{-1} e^{\pm i\om_0 t} f_{\mp}(t) \be_{\mp}(t).
\label{asol}
\eeq
Substitution of this result in Eq.~(\ref{abdot2}) yields
\beq
i{\dot \be}_\pm = \pm {|\bvf(t)|^2 \over \om_0} \be_\pm (t).
\label{bdot}
\eeq
Integrating this equation, we obtain
\beq
\be_\pm (t) = \exp(\mp i\Phi(t)) \be_\pm(0),
\label{bsol}
\eeq
where
\beq
\Phi(t) = \int_0^t dt'\, {|\bvf(t')|^2 \over \om_0}.
\label{Phi}
\eeq
We next note that since by Eq.~(\ref{asol}), $|\al_\pm/\be_\pm| \approx |f_\pm/\om_0| \ll 1$,
$u_\pm \approx \be_\pm$ in leading order in $f_\pm/\om_0$. Equations (\ref{bsol}),
(\ref{Phi}), and (\ref{psispin}) thus give us the desired expression for the time dependence
of a general state.

The above equations also give us the amount of decoherence.
Suppose the initial state of the spin is $2^{-1/2}(\ket+ + \ket-)$, i.e.,
$u_\pm (0) = 2^{-1/2}$. (A product of states of this type for each quantum bit
is the starting point in several quantum algorithms.) Let the state that would be
obtained at time $t$ in the absence of the perturbation  $\bvf$ be $\ket{\psi_0(t)}$.
The states $\ket{\psi_0(t)}$ and $\ket{\psi(t)}$ are
the states of the quantum bit in the ideal and actual QC, without and with decoherence.
The decoherence is given by the overlap
\beq
\langle\psi_0(t) | \psi (t) \rangle = \cos(\Phi(t)).
\label{overlap}
\eeq

Another way to understand these results is that because $|\Bp| \ll B_z$,
the precession axis for the
spin can be taken to be $\hat{\bf z}$ at all times to very good approximation,
and the instantaneous precession frequency can be taken as
\beq
\om'_{0i}  =  (\om_0^2 + V_i^2/\hbar^2)^{1/2} 
           \approx  \om_0 + {V_i^2 \over 2\hbar^2 \om_0}.
\label{ominst}
\eeq
Hence, the time dependence of the states $\ket\pm$ is given by
$\exp\left(\pm i\int_0^t dt'\,\om'_{0i}(t')/2 \right) \ket{\pm}$.
[Note that $\Phi(t)$ is just half the difference between
$\int_0^t \om'_{0i}(t')\,dt'$ and $\om_0t$.]

\subsection{Single and Many Ion Decoherence Times}

The decoherence time for the {\it i\/}th ion is approximately equal to the time required
for its phase to deviate from its unperturbed value by $\pi$: 
$\tau_i \approx \pi/(\om'_{0i} - \om_0)$. Using Eqs.~(\ref{pert}) and (\ref{ominst}),
we obtain
\beq
\tau_i^{-1} \approx {q^2 Q^2 \over 2\pi\hbar^2\om_0}
            \left\langle \biggl(
            \sum_{j\ne i} {u_j \over (z_i -z_j)^4} \biggr) ^2
                                                      \right\rangle ,
\label{taui}
\eeq
where the angular brackets denote some kind of average. In principle this average
should be calculated by expressing the displacements $\bf u$ in terms of the normal
modes of the array, which should be described by a thermal density matrix
at the temperature $T$ to which they have been cooled. (Additional correlations from the
center of mass mode excitations inherent in the two-bit gates should be excluded
since they are part of the intended time evolution and do not generate decoherence.
We have already left this part out in deriving Eq.~(\ref{taui}) by taking $\bvf$
to be slow.) We are interested in obtaining an order of magnitude answer. Therefore,
we will ignore the normal mode structure, and assume each ion to vibrate independently
with some average frequency. Secondly, to obtain the best-case answer, we will assume
that $T \ll \hbar\om_z/k_B$, and simply set $T=0$. To be more concrete, we write
\beq
\langle u_j u_k \rangle = {\hbar \over m \om_t} \dta_{jk},
\label{ujuk}
\eeq
where $\om_t$ is a typical transverse mode frequency, and $m$ is the ionic mass.
It follows that
\beq
\tau_i^{-1} \approx {q^2 Q^2 \over 2\pi\hbar m \om_0\om_t}
                    \sum_{j\ne i}{1\over (z_i - z_j)^8}.
\label{taui2}
\eeq
There are two reasons for using a transverse mode frequency in Eq.~(\ref{ujuk}).
First, in the linear Paul trap, these frequencies are generally higher
than the longitudinal ones. Second, one can arrange via a $J_z$ selection rule
for the longitudinal modes not to excite any $\gtoe$ transitions.

We now combine the individual ion decoherence times to obtain $\tvib$ for the
QC as a whole. The obvious procedure of adding the rates is not quite correct. The reason
can be seen from Eq.~(\ref{overlap}). The overlap between the
actual and intended states of the {\it i\/}th spin is more like
$\cos(t/\tau_i)$ than $\exp(-t/\tau_i)$. Multiplying these overlaps for all spins,
we obtain $P(t)\simeq \prod_i\cos^2(t/\tau_i) \simeq \exp(-t^2/\tvib^2)$ with
\beq
\tvib^{-2} = \sum_i \tau_i^{-2}.
\label{tausq}
\eeq
A more careful justification for this result can be found in Ref.~\cite{agprl}.
We note here that if we simply add $\tau_i^{-1}$ to obtain $\tvib^{-1}$,
we overestimate $\tvib^{-1}$ by a multiplicative factor of
$N^{1/2}$; if this were done, the $N^{1/2}$ in Eq.~(\ref{tvib2}) would
change to $N$.

\subsection{Continuum Approximation for Ion Array}

Equations (\ref{taui2}) and (\ref{tausq}) provide us with a formal answer for the
vibrational decoherence rate. To obtain a more useful result, however, it is necessary
to perform the sums over the lattice positions. For small $N$, say 25 or less, these
sums are best done numerically, but for larger $N$, we can evaluate them by approximating
the ion array as a continuum.

Let us define $s(z_i)$ to be the average of the local spacing between the ion at
$z_i$ and its two nearest neighbours. We expect that for large $N$, this spacing
will vary slowly as we move along the array. Hence, it is a good approximation to
treat $z$ as a continuous variable, and on this basis we will find an expression for
$s(z)$.  We will also find the total length of the array as a function of $N$.

Consider an ion at position $z$, and let us denote the separations to its immediate
neighbours to the left and right by $s_-$ and $s_+$ respectively. The Coulomb force
on the ion at $z$ from these near neighbours is given by
\beq
F^{\rm Coul}_{\rm nn} = q^2(s_-^{-2} - s_+^{-2}) \approx 2q^2s^{-2} (ds/dz),
\label{Fnn}
\eeq
since $s_+ - s_- \approx s(z) (ds/dz)$. For ions not too close to the edge of the
array, the distances to the second, third, fourth, neighbour pairs are approximately
doubled, tripled, quadrupled and so on. The forces from these successively distant
neighbour pairs are smaller than $\Fcnn$ by factors of 4, 9, 16, etc. Since,
$\sum_{n=1}^{\infty} n^{-2} = \pi^2/6$, and since the sum converges very rapidly, we
can write the net Coulomb force on the ion at $z$ as
\beq
F^{\rm Coul}_{\rm net} = (\pi^2q^2/3s^2)(ds/dz).
\label{Fnet}
\eeq
Equating this to the opposing spring force $m\om_z^2 z$ from the trapping potential,
we obtain a differential equation for the spacing function:
\beq
{\pi^2 \over 3 s^2(z)}{ds \over dz} = {z\over d_0^3}.
\label{dsdz}
\eeq
We have introduced
\beq
d_0 = (q^2/m\om_z^2)^{1/3}
\label{d0}
\eeq
as a natural length scale for the trap. (One can easily show that the ion
spacing is of order $d_0$ for 2 or 3 ions in the trap.)

To integrate Eq.~(\ref{dsdz}), we define the total length of the array to be $2L$.
Choosing the center to be at $z=0$, we obtain
\beq
{1\over s(L)} - {1\over s(z)} = - {3\over 2\pi^2d_0^3} (L^2 - z^2).
\label{sin}
\eeq
To make use of this result, however, we need expressions for $L$ and $s(L)$. To find
$s(L)$, we use same argument as was used to obtain Eq.~(\ref{Fnet}), although the
approximation is clearly not as good now. We assume that the spacing between the ions
near the end of the chain is uniform and equal to $s(L)$. The Coulomb force on the
last ion is then $\pi^2 q^2/6 s^2(L)$. Balancing this with the spring force
$m\om_z^2$, we obtain $s(L) \approx \pi (d_0^3/6L)^{1/2}$. Even if this argument is
not watertight, it shows that we can ignore $s^{-1}(L)$ compared to
$3L^2/2\pi^2d_0^3$ in Eq.~(\ref{sin}). This yields
\beq
s(z) = s_0 (1-z^2/L^2)^{-1},
\label{sz}
\eeq
where we have introduced the minimum ion spacing (attained at $z=0$):
\beq
s_0 \equiv s(0) = 2\pi^2d_0^3/L^2.
\label{s0}
\eeq

It still remains to find $L(N)$. If we denote the ion number at position $z$
by $n(z)$, then $dn/dz = 1/s(z)$ in the continuum approximation. Integration of
this result along with Eq.~(\ref{sz}) gives 
\bea
L &=&  d_0(\pi^2 N/2)^{1/3}, \label{LN} \\
s_0 &= & 6.81 d_0 N^{-2/3}. \label{s0N}
\eea
One can also show that the mean spacing varies as $N^{-2/3} \ln N$.

The treatment of the ends of the chain above is not fully satisfactory. Another
continuum approach is due to Dubin.\cite{Dub} He regards the array as fluid of total
charge $qN$. It is known that in a harmonic potential such as that of the trap, such
a fluid forms an ellipsoid of revolution of uniform charge density. When
the trap is much stiffer in the transverse than the longitudinal direction, i.e.,
$\om_t \gg \om_z$, the ellipsoid has a total volume $4\pi N d_0^3$.
Equating the semi major axis of the ellipsoid to $L$, we find the semi minor axis
to be $(3Nd_0^3/L)^{1/2}$. The spacing $s(z)$ is now given by geometry. Let $A(z)$ be
the cross sectional area of the ellipsoid at an axial distance $z$ from the center.
The volume $s(z) A(z)$ clearly contains one unit $q$ of charge, from which it follows
that
\beq
{1\over s(z)} = {3\over 4}{N\over L}\biggl( 1-{z^2 \over L^2} \biggr).
\label{sinDub}
\eeq
This agrees with Eq.~(\ref{sz}), but $s_0$ has a different form. To complete the
solution, we must find this form, or equivalently, $L(N)$. The fluid model answer for $L$
depends on $\om_z/\om_t$, which is clearly wrong if the linear structure is stable.
Dubin therefore resorts to a local density functional theory to estimate the
discreteness correction to the Coulomb energy. He then minimizes the sum of this
correction, the fluid drop self energy, and the trapping potential energy. The resulting
array length is independent of $\om_z/\om_t$ and is given by
\beq
L^3 = 3N \ln(c_0N) d_0^3,
\label{LDub}
\eeq
where $c_0 = 6e^{\gamma - 13/5} \approx 0.8$, and $\gamma$ is Euler's constant.
We also obtain
\beq
s_0 = 4L/3N = 1.92 N^{-2/3} [\ln(0.8 N)]^{1/3} d_0.
\label{sDub}
\eeq
These results differ from Eqs.~(\ref{LN}) and (\ref{s0N}) logarithmically in $N$.

It is interesting to compare these results with some recent semi-numerical work by Meyrath
and James.\cite{mj} They integrate Eq.~(\ref{sz}), write the result as 
$n(z) = az - bz^3$, and fit $a$ and $b$ to power laws in $N$ instead of trying
to relate them to $s_0$ and $L$. They then invert this cubic equation for $z(n)$, and
find that the answer agrees quite well with numerics for $N \ge 25$. Likewise,
Eq.~(\ref{sDub}) provides a very good fit to the numerical results.

\subsection{Lattice Sums}

Equations (\ref{sinDub}-\ref{sDub}) can be used to find the sums in Eqs.~(\ref{taui2})
and (\ref{tausq}). The first type of sum,
\beq
S_n(i) \equiv \sum_{j\ne i}{1\over |z_i - z_j|^n},
\label{sum1}
\eeq
can be very simply evaluated as
\beq
S_n(i) \approx {2 s^n(z_i)} \sum_{j=1}^\infty {1\over j^n}
       = {2\zeta(n) \over s^n(z_i)}.
\label{sumS}
\eeq
This result should hold well for all $i$ except very close to the ends, since the exponent
$n$ is big (3 for an E1 decay, 4 for E2).

The second type of sum is $T_n = \sum_i s^{-n}(z_i)$. Writing $\Delta i \approx dz/s(z)$,
we can approximate it by an integral:
\beq
T_n = \sum_i {1\over s^n(z_i)} \approx \int\limits_{-L}^L {dz\over s^{n+1}(z)}.
\eeq
Substituting Eq.~(\ref{sz}) and performing the integration, we obtain
\beq
T_n \approx {L\over s_0^{n+1}} \biggl( {4\pi \over 4n + 7} \biggr)^{1/2},
\label{sumT}
\eeq
where we have also used an asymptotic formula for $\beta(n+2,1/2)$.

The combination of Eqs.~(\ref{sumS}) and (\ref{sumT}) with Eqs.~(\ref{taui2})
and (\ref{tausq}) leads to the results (\ref{tvib1})-(\ref{tvib2}) quoted in Sec.~2.

\section*{ACKNOWLEDGMENTS}
I am indebted to Daniel James for correspondence, and for sharing the results of his
numerical calculations with me prior to publication.
This work was supported by the National Science Foundation via Grant No. DMR-9306947.

\end{document}